\documentclass[preprint,notoc]{JHEP3}
\usepackage{epsfig}

\def\eslt{E_T^{\rm miss}}

\def\to{\rightarrow}

\def\bi{\begin{itemize}}
 \def\ei{\end{itemize}}
\def\te{\tilde e}

\def\c1p{C1^\prime}
\def\ta{\tilde a}

\def\tu{\tilde u}

\def\ta{\tilde a}

\def\tb{\tilde b}

\def\td{\tilde d}

\def\tst{\tilde t}
\def\ttau{\tilde \tau}

\def\tg{\tilde g}

\def\tw{\widetilde W}
\def\tz{\widetilde Z}
\def\alt{\lesssim}
\def\agt{\gtrsim}
\def\be{\begin{equation}}  
\def\ee{\end{equation}}  
\def\bea{\begin{eqnarray}}  
\def\eea{\end{eqnarray}}  

\def\sps1ap{SPS1a$^\prime$}

\title{Implications of a high mass light MSSM Higgs scalar\\
for SUSY searches at the LHC
}
\author{Howard Baer$^{a}$, Vernon Barger$^b$, Peisi Huang$^b$
and Azar Mustafayev$^{c}$ \\
$^a$Dept.\ of Physics and Astronomy, University of Oklahoma, Norman, OK 73019, USA\\
$^b$Dept. of Physics, University of Wisconsin, Madison, WI 53706, USA\\
$^c$William I. Fine Theoretical Physics Institute, 
University of Minnesota, Minneapolis, MN 55455, USA\\
E-mail: \email{baer@nhn.ou.edu}, \email{barger@pheno.wisc.edu},
\email{phuang7@wisc.edu},\email{mustafayev@physics.umn.edu}}

\preprint{\vbox{UMN--TH--3014/11, FTPI--MINN--11/24}}

\abstract{
The Atlas and CMS collaborations have both reported an excess of events
in the $WW^*\to \ell^+\ell^- +\eslt$ search channel, which could be the
first evidence for the Higgs boson. In the context of the MSSM, the lightest SUSY Higgs scalar
$h$ is expected to occur with mass $m_h\alt 135$~GeV, depending on the range of SUSY
parameters scanned over. Since the $h\to WW^*$ branching fraction falls swiftly
with decreasing $m_h$, a signal in the $WW^*$ channel would favor an $h$ at the
high end of its predicted mass range. We scan over general GUT scale SUSY model parameters
to find those which give rise to $m_h\agt 130$~GeV. 
A value of $m_0\sim 10-20$~TeV is favored, with $A_0\sim \pm 2m_0$,
while the lower range of $m_{1/2}\alt 1$~TeV is also slightly favored. 
This gives rise to an ``effective SUSY'' type of sparticle mass spectrum.
For low $m_{1/2}$, gluino pair production followed by three-body gluino decay to top quarks 
may ultimately be accesible to LHC searches, 
while for higher $m_{1/2}$ values, the SUSY spectra would likely be out of range of 
any conceivable LHC reach.
Since the thermal neutralino relic abundance tends to be very high, late-time entropy dilution or
neutralino decay to light axinos would be required to gain accord with the measured dark matter
abundance.
}  

\begin{document}

\section{Introduction}
\label{sec:intro}

Recent searches for the Standard Model (SM) Higgs boson in the
$H_{SM}\to W^+W^-$ mode have been reported by the Atlas~~\cite{atlas35} and CMS~~\cite{cms35} 
collaborations using 35~pb$^{-1}$ of data, and more recently with 
$1.7$~(1.55)~fb$^{-1}$ of data~~\cite{atlas,cms}. 
The recent analyses allow Atlas and CMS combined to exclude SM-like Higgs bosons in the
mass range 145-288~GeV and 296-466~GeV at 95\% CL. 
Combining these exclusion ranges with the LEP2 limit~~\cite{lep2higgs} that $m_{H_{SM}}>114.4$~GeV, 
we expect the Higgs to inhabit the low mass range 114.4-145~GeV, as expected by precision 
electroweak measurements~~\cite{higgsEW}, or else the Higgs is very heavy.
The low mass Higgs window also
corresponds to the range of $m_h$ expected in the Minimal Supersymmetric Standard Model (MSSM),
where calculations of the lightest, usually SM-like Higgs boson mass $m_h$ require
$m_h\alt 135$~GeV~~\cite{hmass}. 

In addition to exclusion limits, searches in the 
$H_{SM}\to WW^*\to(\ell\bar{\nu}_\ell)+(\bar{\ell}'\nu_{\ell'})$ channel have turned up a roughly 
$2\sigma$ excess by Atlas in events containing no jets, and a similar excess by CMS, but this time in events
containing one jet. 
Also, several events in $H_{SM}\to ZZ^*\to 4\ell$ channel with mass $m(4\ell )\sim 120-140$~GeV have been reported.
As data accrues into the 5-10~fb$^{-1}$ regime, both experiments should gain sensitivity 
to the entire low mass range $m_{H_{SM}}\sim 114-145$~GeV. 
If the present excess of $WW^*$ events persists with an enlarged data set, then these
events will indicate that the Higgs mass exists on the {\it high end} of the low mass window,
since the Higgs branching fraction to $WW^*$ and $ZZ^*$ drops rapidly with decreasing Higgs mass.

In this note, we examine the implications of an $WW^*$ signal in the context of the MSSM,
where the lightest Higgs boson has mass $m_h\alt 135$~GeV. In order to gain a substantial rate for $h\to WW^*$ events, 
we will then expect $m_h\sim 130-135$~GeV in order to maximize the $h\to WW^*$ branching fraction.
By requiring the light Higgs boson $h$ to lie in the 130-135~GeV range, we will find a 
rather tight correlation of model parameters which then offer some rather distinct predictions for 
the nature of superparticle signatures which are also expected at LHC.

\section{Calculations}
\label{sec:calc}

In the MSSM, the Higgs sector consists of two doublet fields $H_u$ and $H_d$, which after the breaking of the
electroweak symmetry, result in the five physical Higgs bosons: two neutral $CP$-even scalars $h$ and $H$, 
a neutral $CP$-odd pseudoscalar $A$ and a pair of charged scalars $H^\pm$~\cite{wss}. 
Over most of the MSSM parameter space the lighest Higgs boson $h$ is nearly SM-like, therefore the SM Higgs search results can
be directly applied to $h$ (for exceptions, see Ref.~\cite{belyaev}). 
A calculation of the light (heavy) scalar Higgs boson mass at 1-loop level 
using the effective potential method gives
\be
m_{h,H}={1\over 2}\left[(m_A^2+M_Z^2+\delta)\mp\xi^{1/2}\right]\, ,
\ee
where $m_A$ is the mass of the $CP$-odd pseudoscalar $A$ and 
\be
\xi = \left[ (m_A^2-M_Z^2)\cos 2\beta +\delta\right]^2+\sin^2 2\beta
(m_A^2+M_Z^2)^2 \, .
\ee
The radiative corrections can be approximated as follows
\be
\delta =\frac{3g^2m_t^4}{16\pi^2M_W^2\sin^2\beta}\log\left[
\left(1+{m_{\tst_L}^2\over m_t^2}\right)\left(1+{m_{\tst_R}^2\over m_t^2}\right)\right] .
\ee
Thus, in order to push the value of $m_h$ to its upper limit, we expect we will have
to probe very large values of top squark soft masses $m_{\tst_{L,R}}$ into the multi-TeV range.

For our calculation of $m_h$, we include the full third generation contribution to 
the effective potential, accounting for all sparticle mixing effects~\cite{bisset}. 
The effective Higgs potential, $V_{eff}$, is 
evaluated with all running parameters in the $\overline{DR}$ renormalization scheme
evaluated at the scale choice $Q_{SUSY}=\sqrt{m_{\tst_1}m_{\tst_2}}$, {\it i.e.} the mean
top squark mass scale. Of particular importance is that the $t$, $b$ and $\tau$
Yukawa couplings are evaluated at the scale $Q_{SUSY}$ using 2-loop MSSM RGEs and 
including full 1-loop MSSM radiative corrections~\cite{pbmz}. 
Evaluating $V_{eff}$ at this (optimized) scale
choice then includes the most important two-loop effects~\cite{hh}. This calculational
procedure has been embedded in the Isajet mass spectra program Isasugra~\cite{isasugra}, 
which we use here for our calculations.

Our first goal is to make a thorough scan of the MSSM model parameter space to search for 
parameter choices leading to the largest values of $m_h$. We will adopt a GUT scale
parameter space for our scan, since this will include the desirable radiative
electroweak symmetry breaking (EWSB) constraint, wherein the large top quark Yukawa coupling $f_t$
plays a crucial role in driving the soft SUSY-breaking parameter $m_{H_u}^2$ to negative values,
so the electroweak symmetry is appropriately broken.
We will also maintain gaugino mass unification, as expected in simple SUSY GUT theories.
However, we will avoid a scan over mSUGRA model parameter space, 
since large values of scalar masses are forbidden beyond the 
hyperbolic branch/focus point (HB/FP) region. Instead, we will scan over the 
two-extra-parameter non-universal Higgs model, dubbed NUHM2~\cite{nuhm}, with parameter choices:
\be
m_0,\ m_{1/2},\ A_0,\ \tan\beta,\ \mu,\ m_A ,
\ee
wherein common soft masses of scalars ($m_0$) and gauginos ($m_{1/2}$) 
along with the common soft trilinear term ($A_0$) are stipulated at the GUT scale, while
the ratio of Higgs vevs ($\tan\beta$), the bilinear superpotential Higgs parameter ($\mu$) and 
the $CP$-odd Higgs mass ($m_A$) are inputted at the SUSY scale $Q_{SUSY}$.

\section{Results}
\label{sec:results}

We employed ISAJET 7.81 to generate 13K random points in the above parameter space, requiring only that
the radiative electroweak symmetry breaking is mantained and the lightest supersymmetric particle (LSP) 
be electrical and color neutral. 
Our scan limits are as follows:
\bea
m_0&:& 0\to 20\ {\rm TeV},\\
m_{1/2}&:& 0\to 5\ {\rm TeV},\\
A_0&:& -5m_0 \to \ +5m_0,\\
\tan\beta&:& 5 \to 55,\\
\mu&:& 0\to 10\ {\rm TeV},\\
m_A&:& 0\to 10\ {\rm TeV}.
\eea
We only scan over positive $\mu$ values so that we do not stray more than
$3\sigma$ away from the measured value of the muon anomalous magnetic moment, $(g-2)_\mu$~\cite{gm2}. We set the mass of the top quark 
$m_t =173.3$~GeV, in accord with the latest Tevatron combination~\cite{top}. 

Our results are shown in Fig.~\ref{fig:scan}, showing 
the dependence of the generated light Higgs mass on each of the model parameters. 
Points satisfying LEP2 chargino bound $m_{\tw_1}>103.5$~GeV~\cite{LEPsusy} are shown as blue dots, 
while those with too low chargino mass $m_{\tw_1}<103.5$~GeV are represented by red crosses. 
The first thing to note is that our scan over parameter
space refines the upper limit on $m_h$ to
\be
m_h<132\ {\rm~GeV} .
\ee
Thus, if $m_h$ comes in much above 132~GeV, then in a SUSY context
we would have to expect some sort of extended Higgs sector, perhaps
the NMSSM~\cite{nmssm} or theories with vector-like matter~\cite{martin}.
We note that we expect just a few~GeV theory error in our $m_h$
calculation.
Also, it should be noted that our value of $m_h$ is typically a couple
GeV below the corresponding FeynHiggs~\cite{feynhiggs} calculation,
mainly due to the fact that we are able to extract and use 
the two-loop $\overline{DR}$ Yukawa couplings including 1-loop
threshold corrections at the $Q_{SUSY}$ scale.
\FIGURE[tbh]{
\includegraphics[width=13cm,clip]{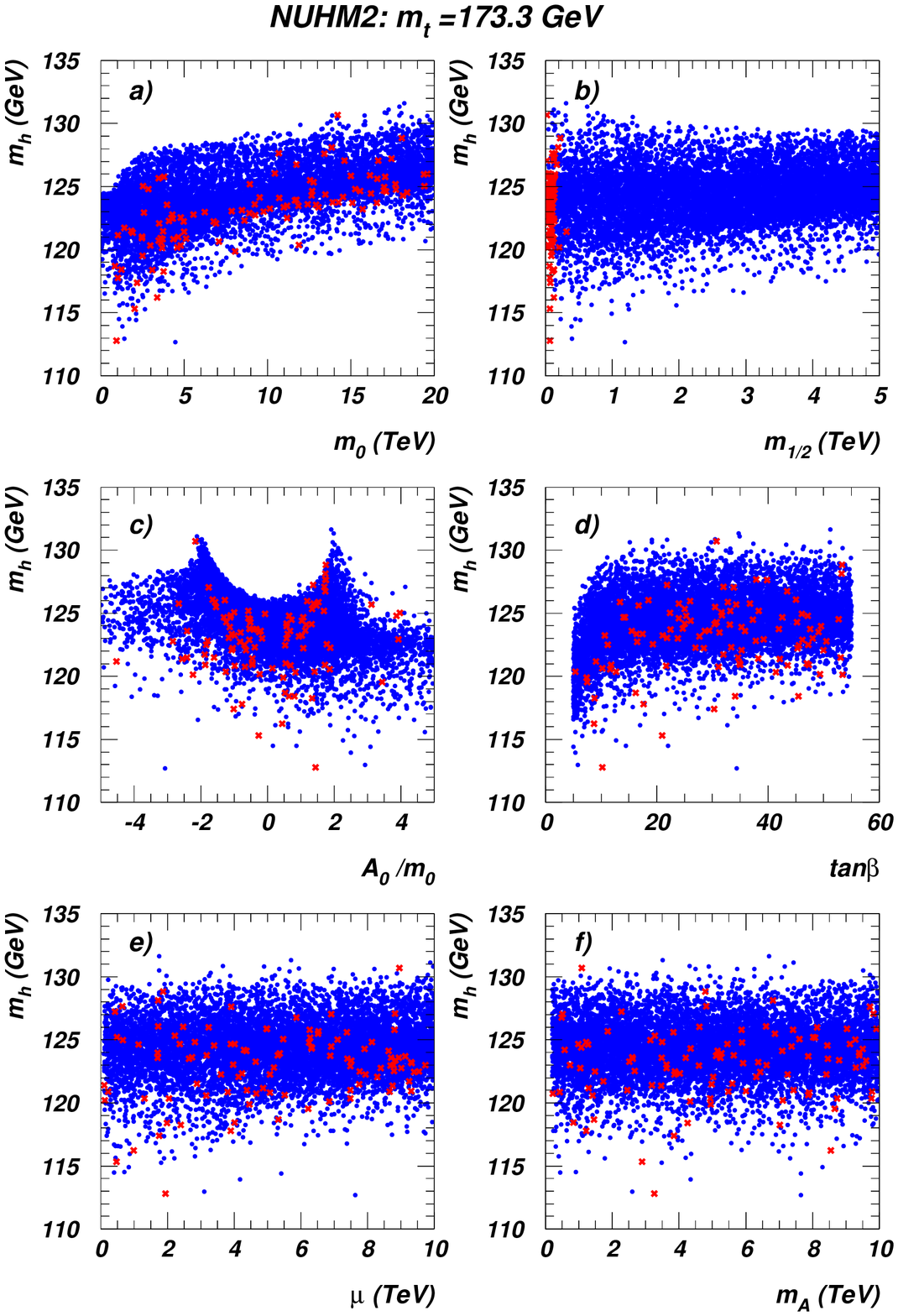}
\caption{A plot of light Higgs mass $m_h$ versus various 
SUSY parameters from a scan over NUHM parameter space.
Red points have charginos masses $m_{\tw_1}<103.5$~GeV, while blue points are
LEP2-allowed.
We take $m_t=173.3$~GeV.
}
\label{fig:scan}}

From frame {\it a}), we see however that $m_h$ can reach to
over 130~GeV only for $m_0$ very high: for values $m_0\agt 10$~TeV.
Thus, if the $h\to WW^*$ signal comes in at very high values
of $m_h\sim 128-132$~GeV, then we can expect squarks and sleptons 
to exist in the multi-TeV regime, well beyond LHC reach.
In addition, since smuons and muon sneutrinos are expected to be multi-TeV, the
value of $(g-2)_\mu$ is expected to be near its SM value.
Alternatively, if $m_h\sim 125$~GeV, then the corresponding
bound on $m_0$ is only $>1$~TeV.

From frame {\it b}), we see that if $m_h\sim 130-132$~GeV, 
then rather low values of $m_{1/2}\alt 1$~TeV are favored, although
some models allow $m_{1/2}$ as high as $2.4$~TeV. 
If the lower portion of the range of $m_{1/2}$ is indeed favored by a heavy Higgs scalar $h$, 
then there may be implications for gluino pair searches at the CERN LHC.
The gluino mass $m_{\tg}$ is shown versus $m_h$ in Fig. \ref{fig:mass}{\it a}).
If we require $m_h\agt 130$~GeV, then we find that $m_{\tg}\alt 4$~TeV, with the region
around $m_{\tg}\sim 1$~TeV being slightly more favored.
In the region of large $m_0$, 
the LHC7 reach~\cite{lhc7} for gluino pair production with 10~fb$^{-1}$ 
is to about $m_{\tg}\sim 800$~GeV, 
while the LHC14 reach~\cite{lhcreach} with 100~fb$^{-1}$ is to $m_{\tg}\sim 1400$~GeV.
Thus, if $m_h\sim 130-132$~GeV, then gluinos might be accessible
to LHC searches, but it is also the case that all sparticles could be 
beyond LHC reach.  
\FIGURE[tbh]{
\includegraphics[width=10cm,clip]{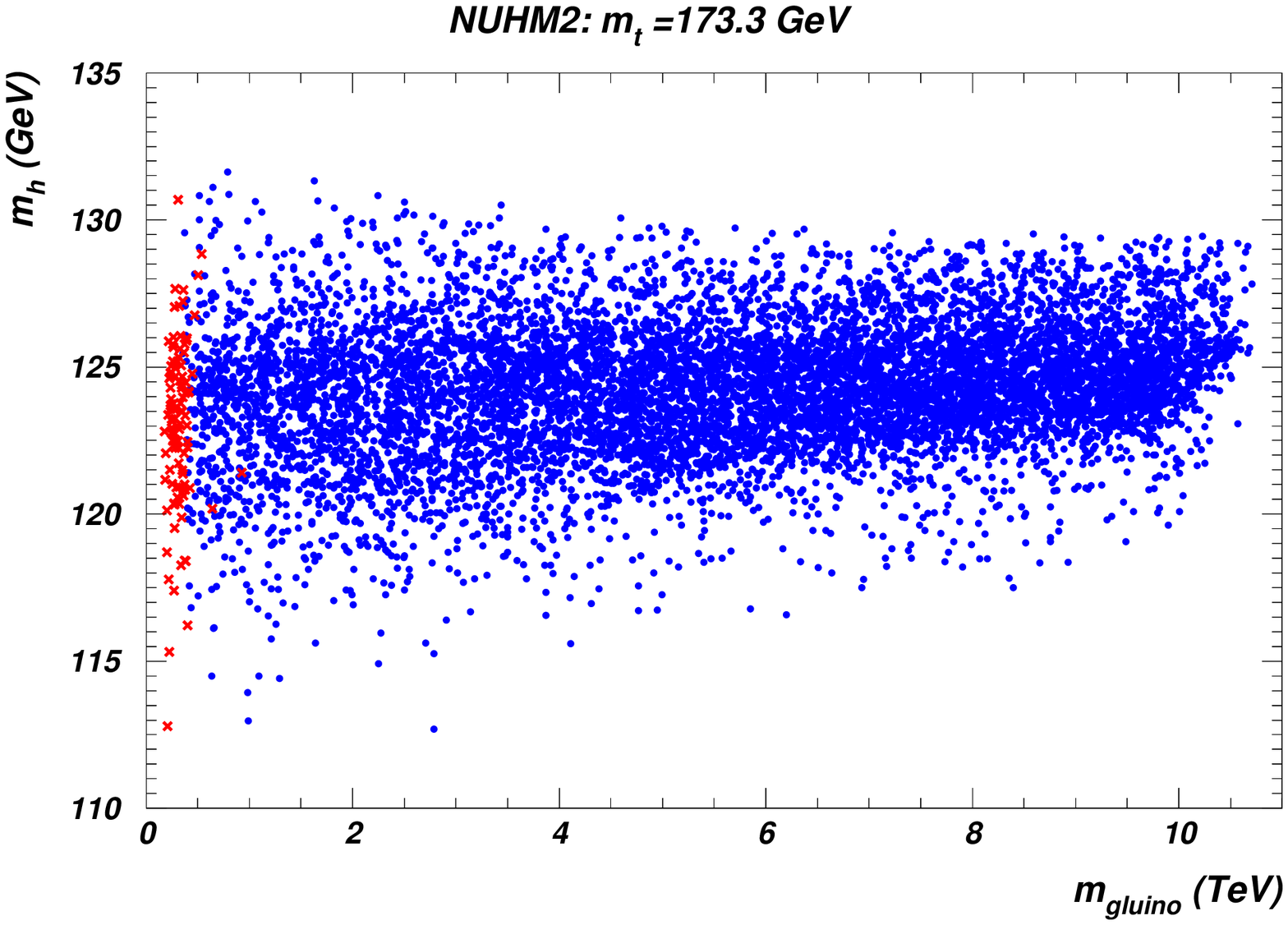}
\includegraphics[width=10cm,clip]{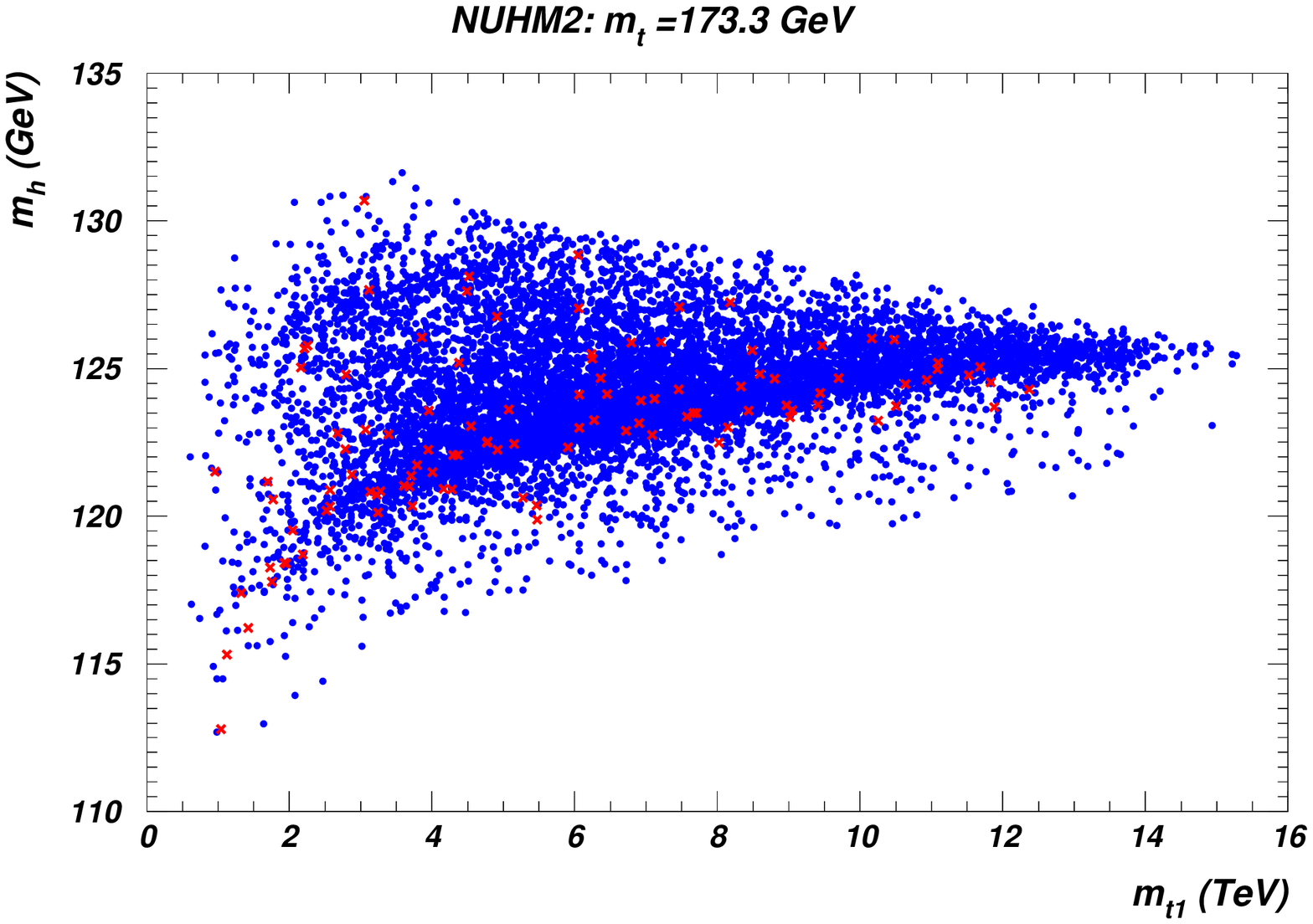}
\caption{A plot of light Higgs mass $m_h$ versus 
$m_{\tg}$ (upper) and $m_{\tst_1}$ (lower) from a scan over NUHM2 parameter space.
We take $m_t=173.3$~GeV.
}
\label{fig:mass}}

In frame {\it c}), we show $m_h\ vs.\ A_0/m_0$. Here we see that the
value of $A_0$ is really restricted to $\sim \pm 2m_0$ in order to attain the very largest
values of $m_h$. For these large $A_0$ values, the top squark mixing is large, which
can suppress the lighter stop mass $m_{\tst_1}$. The mass $m_{\tst_1}$ is shown
versus $m_h$ in Fig.~\ref{fig:mass}{\it b}), where we see that for
$m_h\sim 130-132$~GeV, we have $m_{\tst_1}\sim 2-4$~TeV, even though $m_0$ 
(and hence $m_{\tu,\td}$) is required $\agt 10$ TeV. 
In fact, the boundary conditions of large $m_0$ with low $m_{1/2}$ and $|A_0|\sim 2m_0$
have been derived earlier in the case of Yukawa-unified SUSY~\cite{imh}, wherein third generation
scalar masses are suppressed relative to first/second generation scalars via RG running. These
boundary conditions result in an inverted scalar mass hierarchy (ISMH).

The relatively light top squark mass, along with the large top Yukawa coupling, act to
enhance gluino three-body decays $\tg\to t\bar{t}\tz_i$ (for $i=1-4$)~\cite{btw} at the expense of
three-body decays to first or second generation quarks. Indeed, examining the Isajet sparticle 
decay table for a variety of models with $m_h\sim 130$~GeV shows that $\tg\to t\bar{t}\tz_i$ 
occurs at the 70-80\% level when gluino masses are light enough to be accessible to LHC searches. 

Meanwhile, in frame {\it d}), we see that the largest values
of $m_h$ occur mainly for the upper range of $\tan\beta\sim 15-55$.
From frames {\it e}). and {\it f}), we see that almost any values of
$\mu$ and $m_A$ $\sim .1-10$~TeV are possible if $m_h$ is restricted to be at its upper range.

In Fig.~\ref{fig:bf}, we show the resultant $h\to WW^*$, $ZZ^*$ and $\gamma\gamma$ branching
fractions versus $m_h$ from our scans over NUHM2 parameter space. Indeed, at the very highest
$m_h$ values, we see that $BF(h\to WW^*)\sim20\%$, although it drops by nearly an order of magnitude
as $m_h$ descends into the 110~GeV range. The branching fraction into $ZZ^*$ drops even faster with
decreasing $m_h$, while the $\gamma\gamma$ branching fraction is nearly constant at $\sim 10^{-3}$.
The spread in values comes mainly from the variability in $b$-quark Yukawa coupling due to
its value at $Q_{SUSY}$, which depends on the entire SUSY spectrum via the threshold corrections.
\FIGURE[tbh]{
\includegraphics[width=12cm,clip]{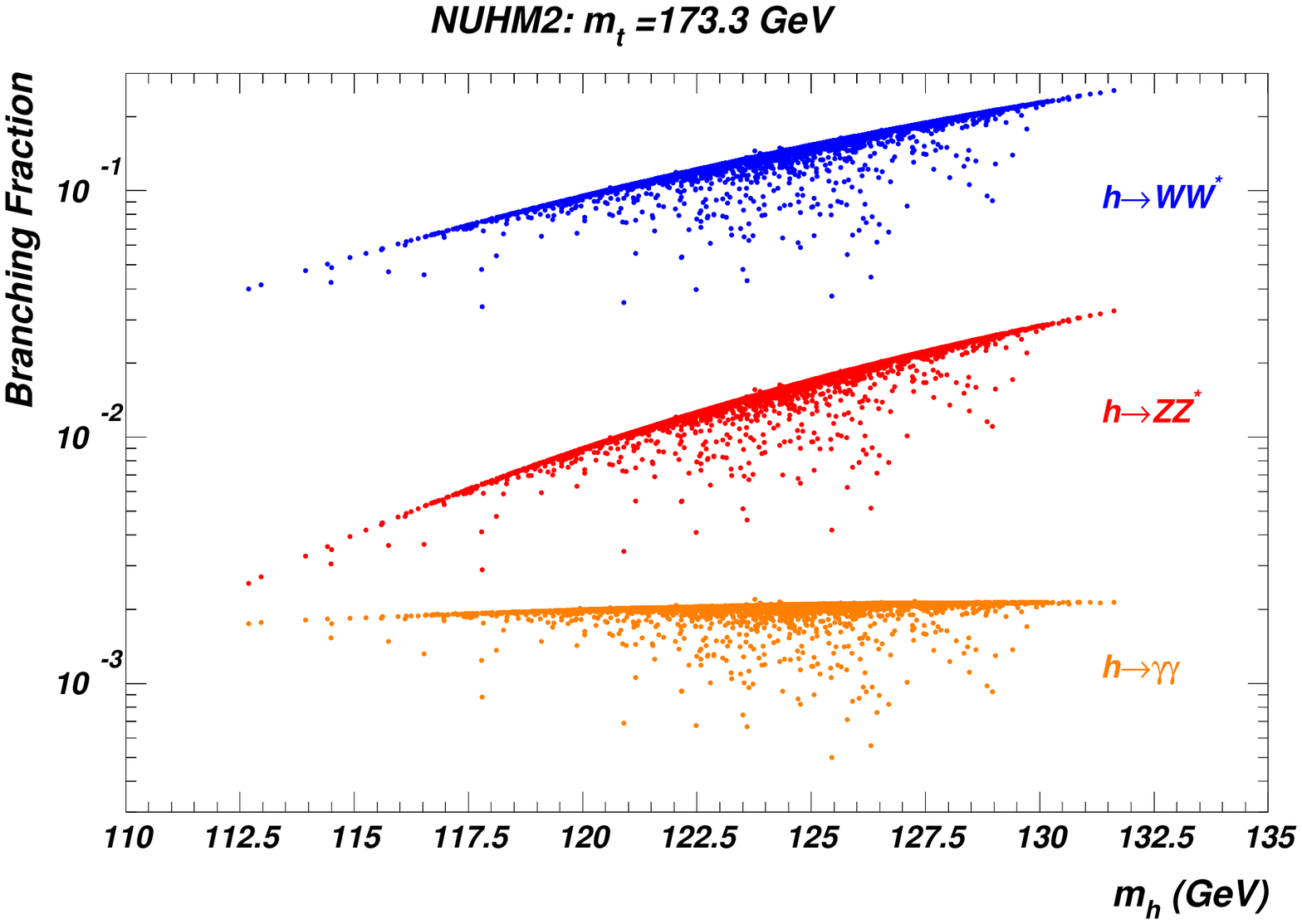}
\caption{A plot of $BF(h\to WW^*)$, $BF(h\to ZZ^*)$ and 
$BF(h\to \gamma\gamma )$ versus $m_h$ from a scan over
NUHM2 parameter space for $m_t=173.3$~GeV.
}
\label{fig:bf}}

In Table~\ref{tab:bm}, we show several sample NUHM2 model points with $m_h\sim 130$~GeV. The first
two points, one for $A_0>0$ and one for $A_0<0$, have gluino masses within reach of LHC with
dominant $\tg\to t\bar{t}\tz_1$ decay, so we would expect LHC collider events containing
four top quarks plus $\eslt$ from the escaping neutralinos. The third point has $m_{\tg}\sim 1800$~GeV.
In this case, we would expect LHC to see a light Higgs scalar with $m_h\sim 130$~GeV, but little or 
no sign of supersymmetry: the SUSY spectrum essentially decouples from LHC searches due to too heavy 
a mass spectrum.

We also show in Table~\ref{tab:bm} the standard thermal neutralino dark matter abundance, assuming
neutralino-only dark matter, as calculated by IsaReD~\cite{isared}. We see that 
$\Omega_{\tz_1}^{std}h^2$ is typically 4-5 orders of magnitude larger than the WMAP-measured value~\cite{wmap7}
of $\Omega_{CDM}h^2=0.1123\pm 0.0035$ (68\% CL), 
so that under a standard cosmology, these points would be excluded.
This is similar to what occurs in Yukawa-unified of ``effective SUSY'' models~\cite{esusy}, where a spectrum of 
lighter gauginos plus multi-TeV scalars results in a standard dark matter abundance which is several
orders of magnitude beyond observation. There are several appealing ways around this situation.
In one case, one may postulate the existence of additional scalar fields with mass in the 10-100~TeV
range and with delayed decays, which occur shortly before BBN begins. 
This could be the case for instance for light moduli fields of string theory~\cite{lightmod}, or for saxions
from a Peccei-Quinn (PQ) solution to the strong CP problem~\cite{pqww} (or both). 
In these cases, the scalar fields, 
which can be produced typically via coherent oscillations, can inject considerable entropy into the early
universe, thus diluting all relics present at the time of decay~\cite{mr,gg,AY,bigfa}.
A more conservative possibility is to choose NUHM2 model parameters with very low $\mu$ values such that
the lightest neutralino is mixed- or mainly- higgsino-like, or to choose $m_A$ values such
that $m_A\sim 2m_{\tz_1}$ so that neutralinos annihilation is enhanced via the $A$-resonance\cite{nuhm}.

Another possibility also occurs in the PQ augmented MSSM, where the $R$-parity odd axinos $\ta$ are
the lightest SUSY particles, and at the MeV scale. In this case, the thermally produced neutralinos
would decay via $\tz_1\to\ta\gamma$ with lifetimes of order $\alt 1$~sec, so that the 
(non-thermally produced, NTP) axino abundance is 
$\Omega_{\ta}^{NTP}h^2=\frac{m_{\ta}}{m_{\tz_1}}\Omega_{\tz_1}^{std}h^2$~\cite{ckkr}. Since the factor 
$\frac{m_{\ta}}{m_{\tz_1}}\sim 10^{-5}$, the neutralino overabundance is ultimately erased.
The remaining dark matter fraction may be built up from a combination of thermally produced
axinos~\cite{th_axinos}, along with axions produced via vacuum misalignment~\cite{absik,bbs}.

\begin{table}
\begin{center}
\begin{tabular}{lccc}
\hline
parameter & Pt. 1 & Pt. 2 & Pt. 3 \\
\hline
$m_0 \ [{\rm TeV}]$  & 18.2 & 17.246 & 14.169 \\
$m_{1/2}\ [{\rm TeV}]$   & 307.8 & 122.58 & 712.74 \\
$A_0\ [{\rm TeV}]$ & 34.737 & -36.576 & 28.588 \\
$\tan\beta$ & 51.19 & 34.9 & 43.3 \\
$\mu\ [{\rm TeV}]$       & 1.759 & 9.880 & 5.660 \\
$m_A\ [{\rm TeV}]$       & 6.695 & 7.435 & 2.189 \\
\hline
$m_h\ [{\rm~GeV}]$       & 131.62 & 131.1 & 130.4 \\
$m_{\tg}\ [{\rm~GeV}]$   & 789.0 & 648.3 & 1825.3 \\
$m_{\tu_L}\ [{\rm TeV}]$ & 18.117 & 17.139 & 14.121 \\
$m_{\te_L}\ [{\rm TeV}]$ & 18.218 & 17.326 & 14.285 \\
$m_{\tst_1}\ [{\rm TeV}]$& 3.581 & 3.766 & 2.957 \\
$m_{\tb_1}\ [{\rm TeV}]$ & 9.434 & 9.763 & 8.418  \\
$m_{\ttau_1}\ [{\rm TeV}]$ & 8.147 & 13.078  & 8.926 \\
$m_{\tw_1}\ [{\rm~GeV}]$ & 176.0 & 201.1 & 568.9 \\
$m_{\tz_2}\ [{\rm~GeV}]$ & 174.8 & 198.8 & 565.2 \\
$m_{\tz_1}\ [{\rm~GeV}]$ & 92.4 & 93.9 & 292.2 \\
\hline
$\Omega_{\tz_1}^{std}h^2$ & $8.3\times 10^3$ & $1.7\times 10^4$ & $1.1\times 10^3$
\\

\hline
\end{tabular}
\caption{Masses and parameters in~GeV/TeV units
for several high $m_h$ NUHM2 SUSY
models using Isajet~7.82 with $m_t=173.3$~GeV.
}
\label{tab:bm}
\end{center}
\end{table}

\section{Conclusions}
\label{sec:conclude}

The recent surplus of $WW^*$ events above the SM background, as measured by both Atlas and CMS experiments, 
may point to a light MSSM Higgs scalar boson $h$ at the upper edge of its expected mass 
range: $m_h\sim 128-132$~GeV. We have scanned over NUHM2 model parameter space, which maintains
the desirable feature of radiative EWSB, while allowing for scalar masses beyond
the HB/FP limit: $m_0\sim 5-20$ TeV. By requiring $m_h\agt 128$~GeV, we find
that $m_0\sim 10-20$ TeV is required, with $|A_0|\sim 2m_0$. While a wide range of
$\tan\beta$, $\mu $ and $m_A$ values are allowed, the value of $m_{1/2}$ has a mild preference
for the low end of its range. 

The associated SUSY particle spectra turns out to be of the ``effective SUSY'' type~\cite{esusy},
with multi-TeV first/second generation scalars, few-TeV third generation scalars
and possibly sub-TeV gauginos. In this case, SUSY signatures at LHC should be
dominated by gluino pair production, with dominant $\tg\to t\bar{t}\tz_i$ decays: thus, a 
corroborating signal would be in the $4t+\eslt$ channel. 
It is also possible that the entire SUSY spectrum is quite heavy, and beyond LHC reach.
The thermal neutralino dark matter  abundance is predicted to be far above the WMAP7 
measured value (unless very low $\mu$ or $m_A\sim 2m_{\tz_1}$ is chosen), 
so that a diminution of neutralinos either via late-time entropy injection 
or by decays to
MeV-scale axinos would be needed to reconcile with the measured dark matter abundance.

\acknowledgments

This work was supported in part by the U.S. Department of Energy under grants DE-FG02-04ER41305, 
DE-FG02-95ER40896 and DE--FG02--94ER--40823.


%

\end{document}